\documentclass[prl, superscriptaddress, floatfix,twocolumn,showpacs,amsfonts,amsmath,amssymb,final]{revtex4}

\usepackage{graphicx}

\usepackage{epstopdf}
\usepackage[countmax]{subfloat}

\usepackage{dcolumn} 

\usepackage{bm} 

\usepackage{color}

\usepackage{float}

\usepackage{latexsym}

\usepackage{feynmf}

\newcommand {\etalc}{{\it et al}., }

\begin{document}

\title{Spin liquid phase due to competing classical orders in the semiclassical theory of the Heisenberg model with ring exchange on an anisotropic triangular lattice}

\author{Michael Holt}

\affiliation{Centre for Organic Photonics and Electronics, School of Mathematics \& Physics, University of Queensland, Brisbane, Queensland 4072, Australia}

\author{Ben. J. Powell}

\affiliation{Centre for Organic Photonics and Electronics, School of Mathematics \& Physics, University of Queensland, Brisbane, Queensland 4072, Australia}

\author{Jaime Merino}

\affiliation{Departamento de F\'isica Te\'orica de la Materia Condensada, Condensed Matter Physics Center (IFIMAC) and Instituto Nicol\'as Cabrera, Universidad Aut\'onoma de Madrid, Madrid 28049, Spain}


\begin{abstract}

Linear spin wave theory shows that ring exchange induces a quantum disordered region in the phase diagram of the title model. Spin wave spectra show that this is a direct manifestation of competing classical orders. A spin liquid is found in the `Goldilocks zone' of frustration, where the quantum fluctuations are large enough to cause strong competition between different classical orderings but not  strong enough to stabilize spiral order. We note that the spin liquid phases of $\kappa$-(BEDT-TTF)${_2}X$ and $Y$[Pd(dmit)$_2$]$_2$ are found in this Goldilocks zone.

\end{abstract}

\maketitle

Quantum spin liquids are characterized by ground states with no long-range magnetic order and no breaking of spatial (rotational or translational) symmetries that are not adiabatically connected to the band insulator \cite{Balents10, Normand09, Powell11}. Recently a number of experiments have identified a handful of materials as candidate spin liquids \cite{Lee08, Shimizu03, Yamashita08, Yamashita09, Yamashita10, Itou10, Kanoda10}. That these candidate spin liquids were only found after many decades of searching already hints that conditions must be just right for a quantum spin liquid to emerge.

Here we focus on Mott insulating phases of two related families of organic charge transfer salts: $\kappa$-(BEDT-TTF)$_2X$ and $Y$[Pd(dmit)$_2$]$_2$, where $X$ and $Y$ are (typically inorganic) counter-ions. Each family includes a candidate spin liquid: $\kappa$-(BEDT-TTF)$_2$Cu$_2$(CN)$_3$ \cite{Shimizu03, Yamashita08, Yamashita09, Kanoda10} and Me$_3$EtSb[Pd(dmit)$_2$]$_2$ \cite{Yamashita10, Itou10, Kanoda10}, where Et =  C$_{2}$H$_{5}$ and Me =  CH$_3$. But other members of each family display long range magnetic order, for example, $X=$Cu[N(CN)$_2$]Cl or Cu[N(CN)$_2$]Br (for deuterated BEDT-TTF) and $Y=$Me$_4$P, Me$_4$As, EtMe$_3$As, Et$_2$Me$_2$P,  Et$_2$Me$_2$As and Me$_4$Sb \cite{Shimizu07, Powell11, Kanoda10}. 

The key question then is: what is the physics that determines whether the ground state is magnetically ordered or not?  In this Letter we study the Heisenberg model on the ATL with ring exchange using linear spin wave theory (LSWT). We show that in weakly frustrated systems long-range magnetic order is robust to ring exchange. At intermediate frustration, the quantum fluctuations induced by ring exchange suppress long range magnetic order while in strongly frustrated systems where fluctuations become more important long-range spiral order persists in the presence of ring exchange. This is highly analogous to the `order-by-disorder' mechanism due to quantum or thermal fluctuations \cite{Villain, Chandra90, Chubukov92}. Therefore, we argue that there is a Goldilocks zone of frustration, where quantum fluctuations are large enough to cause strong competition between different classical orderings but not so strong to stabalise spiral order. This is entirely consistent with recent electronic structure calculations that show that organic charge transfer salts with spin liquid ground states correspond to the intermediate frustration of the Goldilocks zone \cite{Scriven12, Kandpal09, Nakamura09, Nakamura12, Tsumuraya13}. Analysis of the spin wave spectra show that the spin liquid state is a consequence of competition between classical ordered states. Thus we conclude that the interplay of ring exchange and geometrical frustration is responsible for the spin liquid state found. Our results are relevant to weak Mott insulators {\it i.e.} insulators lying close to the insulator-to-metal transition so that ring exchange is relevant.

\begin{figure}
 \begin{center}
     \includegraphics[trim = 5mm 5mm 5mm 5mm, width=0.95\columnwidth,clip]{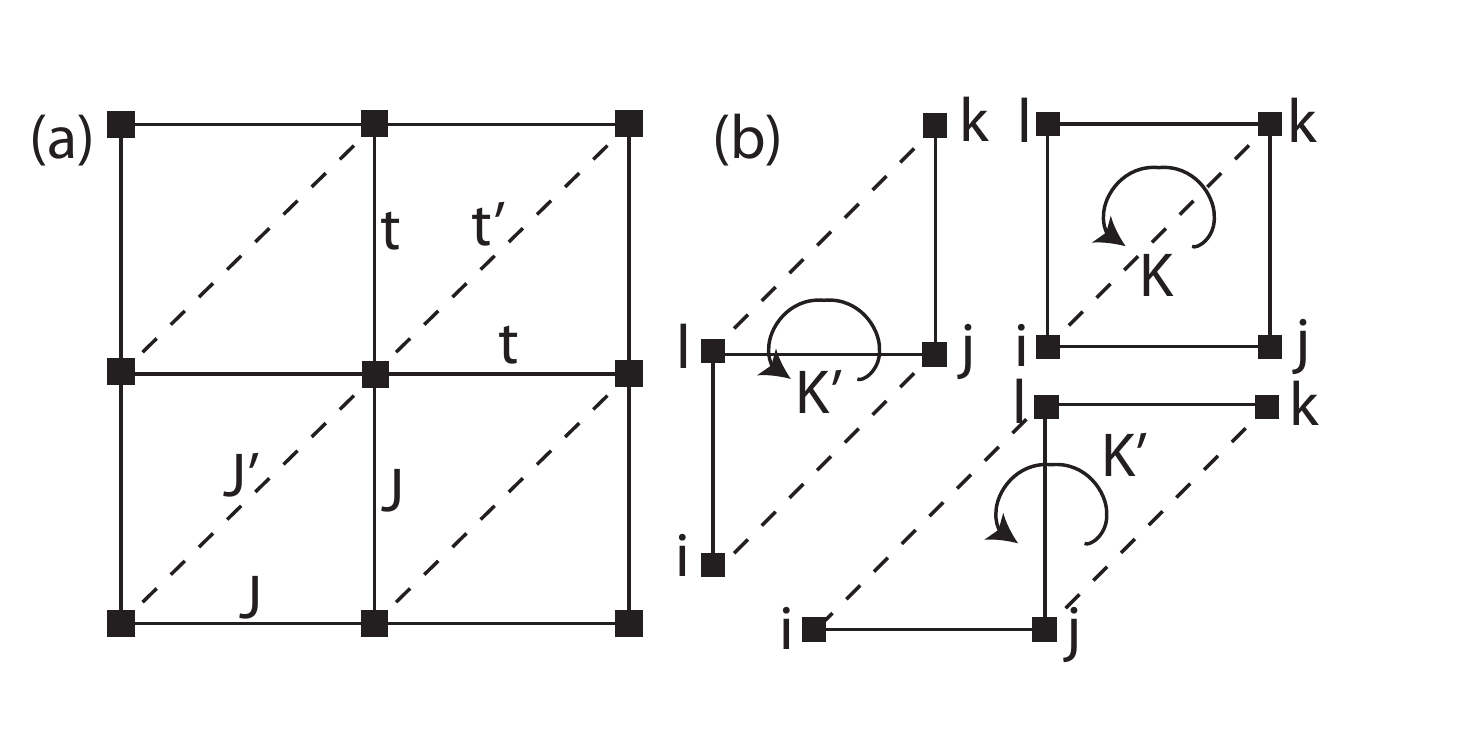}
    \caption{(a) Sketch of the ATL showing the exchange interactions $J$ and $J'$, which act in the nearest-neighbor and next-nearest neighbor along one diagonal directions as shown.  (b) The three distinct ways to draw the four-site plaquettes relevant to ring exchange on the ATL.}
    \label{fig:schematic}
    \end{center}
\end{figure}

The simplest model for the insulating states of the $\kappa$-(BEDT-TTF)$_2X$ and $Y$[Pd(dmit)$_2$]$_2$ salts is the half-filled Hubbard model on the anisotropic triangular lattice (ATL) (Fig.\ref{fig:schematic}a) \cite{Powell11}, where each site represents a dimer, (BEDT-TTF)$_2$ or [Pd(dmit)$_2$]$_2$. The ATL is also realized in Cs$_2$CuCl$_4$ with $J'/J\approx 3$ \cite{Fjaerestad07}, however, the effect of ring exchange is expected to be smaller in this material. This model contains three parameters: $U$ the effective on-site Coulomb repulsion, $t$ the nearest neighbor hopping integral and $t'$ the next nearest neighbor hopping integral along one diagonal only. For $U\gg t,t'$, i.e., deep into the Mott insulating phase, the model simplifies further to the Heisenberg model on the ATL with $J=4t^2/U$ and $J'=4t'^2/U$ to leading order. Electronic structure calculations \cite{Scriven12} suggest that both spin liquids and the valence bond-solid, Me$_3$EtP[Pd(dmit)$_2$]$_2$, have
  $0.5\lesssim J'/J\lesssim0.8$; whereas salts that display long range order have either  $J'/J\lesssim0.5$ or $J'/J\gtrsim 0.8$.

Anderson first proposed the resonating valence bond (RVB) spin liquid state as a possible ground state of the isotropic triangular lattice ($J'=J$) \cite{Anderson74}. However, later numerical work \cite{Sindzingre94} has shown that the ground state has `$120^\circ$ order' - a special case of  spiral order, discussed below, with an ordering wavevector ${\bf Q}=(2\pi/3,\pi/3)$. A range of other methods have been used to study the ATL Heisenberg model including linear spin wave theory \cite{Merino99, Trumper99}, series expansions \cite{Fjaerestad07, Zheng99}, the coupled cluster method \cite{Bishop09}, large-N expansions \cite{ChungJPCM01}, variational Monte Carlo \cite{Monte}, resonating valence bond theory \cite{RVB, RVB1, RVB2, RVB3, RVB4}, pseudo-fermion functional renormalization group \cite{Reuther11}, slave rotor theory \cite{Rau11}, renormalisation group \cite{Starykh07}, and the density matrix renormalisation group \cite{Weng06}. These calculations show that for weak frustration N\'eel $(\pi, \pi)$ order is realised and spiral $(q,q)$ long range AFM order is realized for $J'/J\sim1$. There remains controversy as to whether another state is realized between these two phases, but no conclusive evidence for a spin liquid ground state has been found in this model. 

Many of the organic charge transfer salts considered here undergo Mott metal-to-insulator transitions under relatively modest hydrostatic pressures \cite{Powell06, Kanoda10}. This suggests that higher order terms in the $U/t$ expansion may be relevant. Furthermore, there is significant variation in the critical pressure required to drive the Mott transition in different salts \cite{Yamaura04} which suggests that different salts represent different values of $U/t$ and not just different values of $t'/t$. There has been far less investigation of how this affects the properties of the materials. If one continues to integrate out the charge degrees of freedom, the first non-trivial new terms appear at fourth order with the `ring-exchange' processes illustrated in Fig. \ref{fig:schematic}b [see also Eq. (\ref{eq:bigHamiltonian}), below]. There are two distinct ring exchange term on the ATL: $K=80t^4/U^3$ and $K'=80t^2t'^2/U^3$ to lowest order \cite{MacDonald88, Balents03}. Note that the large prefactor means that the ring exchange term is relevant to larger values of $U/t$ than one would expect na\"ively. It has been argued \cite{Motrunich05, Yang10} near the Mott transition ring exchange destroys the long range magnetic order. In particular, for $J'=J$ and $K'=K$ Motrunich \cite{Motrunich05} found that AFM order is preserved for small $K/J \lesssim 0.14 - 0.20$ \cite{fn1} but is destroyed for larger $K/J$ leading to a gapped spin liquid for $K/J > 0.28$. However, this that applying pressure, which decreases $U/t$, should drive a magnetically ordered to spin liquid transition; which has not been observed in the antiferromagnetically ordered organic charge transfer salts with $t'\simeq t$.

The only work we are aware of to discuss the Heisenberg model with ring exchange on the ATL consider two leg \cite{Sheng09} and four leg \cite{Block11} ladders. Both studies suggest the existence of quantum spin liquids. Therefore, it is important to ask how these states survive as one moves to the full two-dimensional problem.   

Hauke \cite{Hauke13} has considered a model with three distinct exchange interactions and argued that this can explain the phase diagram of the organic charge transfer salts. Alternative models such as the quarter-filled Hubbard model, with each site representing a monomer \cite{Li10} and multi-orbital models \cite{Nakamura12} have also been proposed.

We consider the multiple-spin exchange hamiltonian \cite{Thouless65, Misguich99} on an ATL involving ring exchange on four sites:
\begin{eqnarray}
\label{eq:bigHamiltonian}
\hat{H} &=& \bigg (\frac{J}{2}\sum_{
\begin{picture}(11,11)(0,0)
	\put (0,9) {\line (1,0) {10}}
	\put (0,9) {\circle*{4}}
	\put (10,9) {\circle*{4}}
\end{picture}} + \frac{J}{2}\sum_{
\begin{picture}(11,11)(0,0)
	\put (5,0) {\line (0,1) {10}}
	\put (5,0) {\circle*{4}}
	\put (5,10) {\circle*{4}}
\end{picture}} + \frac{J}{2}'\sum_{
\begin{picture}(11,11)(0,0)
	\put (0,0) {\line (1,1) {10}}
	\put (0,0) {\circle*{4}}
	\put (10,10) {\circle*{4}}
\end{picture}}\bigg )\hat{P}_{ij}\nonumber\\ 
&& +  \bigg( K \sum_{\begin{picture}(11,11)(0,0)
 \put (0,0) {\line (1,0) {10}}
        \put (0,10) {\line (1,0) {10}}
        \put (0,0) {\line (0,1) {10}}
        \put (10,0) {\line (0,1) {10}}
        \put (0,10) {\circle*{4}}
        \put (10,10) {\circle*{4}}
        \put (0,0) {\circle*{4}}
        \put (10,0) {\circle*{4}}
\end{picture}}
+   K'\sum_{\begin{picture}(11,11)(0,0)
 \put (-5,0) {\line (1,0) {10}}
        \put (5,10) {\line (1,0) {10}}
        \put (-5,0) {\line (1,1) {10}}
        \put (5,0) {\line (1,1) {10}}
        \put (-5,0) {\circle*{4}}
        \put (15,10) {\circle*{4}}
        \put (5,0) {\circle*{4}}
        \put (5,10) {\circle*{4}}
\end{picture}}
+ K'\sum_{\begin{picture}(11,11)(0,0)
 \put (0,-10) {\line (0,1) {10}}
        \put (0,-10) {\line (1,1) {10}}
        \put (0,0) {\line (1,1) {10}}
        \put (10,0) {\line (0,1) {10}}
        \put (10,10) {\circle*{4}}
        \put (10,0) {\circle*{4}}
        \put (0,0) {\circle*{4}}
        \put (0,-10) {\circle*{4}}
\end{picture}}
\bigg)
\bigg (\hat{P}_{ijkl} + \hat{P}_{lkji}\bigg )
\end{eqnarray}
\vspace*{2pt}

\noindent where $\hat{P}_{ij} = 2\hat{\bf S}_{i} \cdot \hat{\bf S}_{j} + 1/2$ permutes the spins on sites $i$ and $j$,  $\hat{\bf S}_{i}$ is the usual spin operator on site $i$,  and $\hat{P}_{ijkl}=\hat{P}_{ij}\hat{P}_{jk}\hat{P}_{kl}$   cyclically permutes the four spins around the four-site plaquettes, cf. Fig. \ref{fig:schematic}b.  At this point it is helpful to note that, to lowest order in $t/U$ and $t'/U$, $K'/K = J'/J$. In this Letter we take this equality to hold, primarily to limit the size of the parameter space of the model.

\begin{figure}
 \begin{center}
     \includegraphics[trim = 0mm 20mm 0mm 0mm, width=0.98\columnwidth, clip]{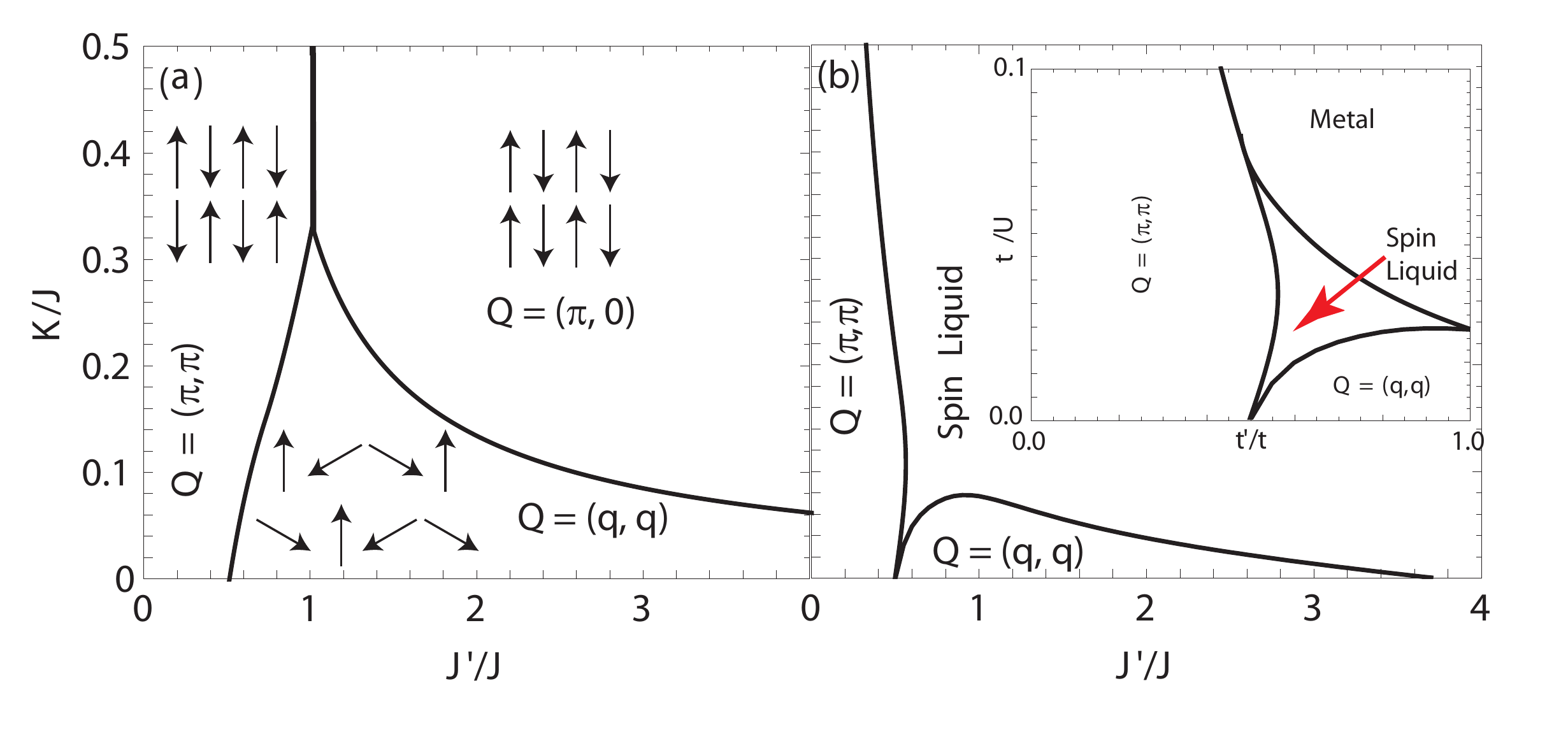}
    \end{center}
    \caption{(a) Classical and (b) quantum (LSWT) phase diagrams for the ATL with ring exchange. It is clear that, even in these semiclassical calculations, quantum fluctuations strongly suppress long-range order when ring exchange is introduced. In panel (a) we also show cartoons of the magnetic orders found to be stable. The inset to panel (b) shows a sketch of a proposed phase diagram for the Hubbard model on the ATL based on the calculation reported here.}
    \label{fig:phasering}
\end{figure}

We plot the classical phase diagram in Fig \ref{fig:phasering}a. We find N\'{e}el, collinear, and spiral phases with ordering vectors  ${\bf Q} = (\pi, \pi)$, ${\bf Q} = (\pi, 0)$, and  ${\bf Q} = (q, q)$ with $q = \arccos( -[2J' + 2K + 8K' - \{4(J' + K + 4K')^2 - 48(J + 2K)K'\}^{1/2}]/24K' )$ respectively \cite{fncpd}. For $K = 0$ we observe a transition from N\'{e}el to spiral order for $J'/J = 0.5$, consistent with previous studies of the Heisenberg model \cite{Merino99, Trumper99}. With increasing ring exchange the N\'{e}el and collinear phases are stabilized, while the spiral order is destabilized. Even at the classical level the spiral phase is most stable to ring exchange when $J' = J$, \emph{i.e.}, when the system is most strongly frustrated.  We will see below that this stabilisation of the spiral phase is reflected in the quantum calculations.

We study the quantum phase diagram and elementary excitations for $S = 1/2$ at $T = 0$ using LSWT. It is  convenient \cite{Miyake92, Singh92}  to rotate the quantum projection axis of the spins at each site along its classical direction $\hat{S}_{i}^{x'} = \tilde{S}_{i}^{x}\cos(\theta_{i}) + \tilde{S}_{i}^{z}\sin(\theta_{i})$, $\hat{S}_{i}^{y'} = \tilde{S}_{i}^{y}$, $\hat{S}_{i}^{z'} = -\tilde{S}_{i}^{x}\sin(\theta_{i}) + \tilde{S}_{i}^{z}\cos(\theta_{i})$ where $\theta_{i} = {\bf Q} \cdot {\bf r}_{i}$, where ${\bf r}_{i}$ is the  position of the $i^\textrm{th}$ spin. This  simplifies the spin-wave treatment with the result that only one, rather than three, species of boson is required to describe the spin operators. The bosonization of the spin operators is performed via the Holstein-Primakov transformation $\tilde{S}_{i}^{z} = S - \hat{a}_{i}^{\dagger} \hat{a}_{i}$, $\tilde{S}_{i}^{+} = \sqrt{2S - \hat{a}_{i}^{\dagger} \hat{a}_{i}}\hat{a}_{i}$,  $\tilde{S}_{i}^{-} = \hat{a}
 _{i}^{\dagger}\sqrt{2S - \hat{a}_{i}^{\dagger} \hat{a}_{i}}$, where $\tilde{S}_{i}^{\pm} = \tilde{S}_{i}^{x} \pm i \tilde{S}_{i}^{y}$.  LSWT   takes the leading order terms in a $1/S$ expansion, which describe noninteracting spin waves. At this level of approximation ring-exchange contributes by dressing the effective two spin exchange and, in particular, introduces additional long-range frustrated interactions. We proceed by diagonalizing the Fourier transformed Hamiltonian via a Bogoliubov transformation, which yields
\begin{equation}
\label{eq:diagonalizedHamiltonian}
\hat{H} = E_{GS}^{(0)} + \frac{1}{2}\sum_{\bf k} (\omega_{\bf k} - A_{\bf k}) + \sum_{\bf k} \omega_{\bf k} \hat{\alpha}_{\bf k}^{\dagger} \hat{\alpha}_{\bf k},
\end{equation} 
\noindent where $E_{GS}^{(0)}$ is classical ground state energy, $\omega_{\bf k} = \sqrt{A_{\bf k}^2 - 4B_{\bf k}^2}$ is the spin-excitation spectrum, $A_{\bf k} = \frac{1}{4} [J_{{\bf Q}+{\bf k}} + J_{{\bf Q}-{\bf k}} ] + J_{\bf k}/2 - J_{\bf Q}$,  $B_{\bf k} = \frac{1}{8} [J_{{\bf Q}+{\bf k}} + J_{{\bf Q}-{\bf k}}] - J_{\bf k}/2$, and $J_{\bf k}$ is the Fourier transform of the exchange interaction \cite{fnJ}.

The LSWT phase diagram is shown in Fig. \ref{fig:phasering}(b). In studying the quantum phase diagram  it is key to consider the staggered magnetization $m_{s} =\langle \tilde{S}_{i}^{z} \rangle = S + 1/2 - \int_{BZ}d^2k A_{\bf k}/8\pi^2\omega_{\bf k}$ as the vanishing of of $m_{s}$ indicates a quantum disordered state. The most striking feature of the phase diagram is that, even in this semi-classical theory, quantum fluctuations destroy long range magnetic order over large areas of the phase diagram. These quantum disordered regions occur in the parameter region consistent with DMRG calculations on four-leg triangular ladders \cite{Block11}.  

Examination of both the ground state energy and the staggered magnetization indicates that both the N\'{e}el-spin liquid and spiral-spin liquid phase boundaries are lines of first order phase transitions vanishing at a quantum critical point at $K=K'=0$, $J'/J=0.5$. This is consistent with what has previously been found in the $K=K'=0$ case \cite{Merino99,Trumper99}.

\begin{figure*}
 \begin{center}
 \includegraphics[width=0.66\columnwidth, clip]{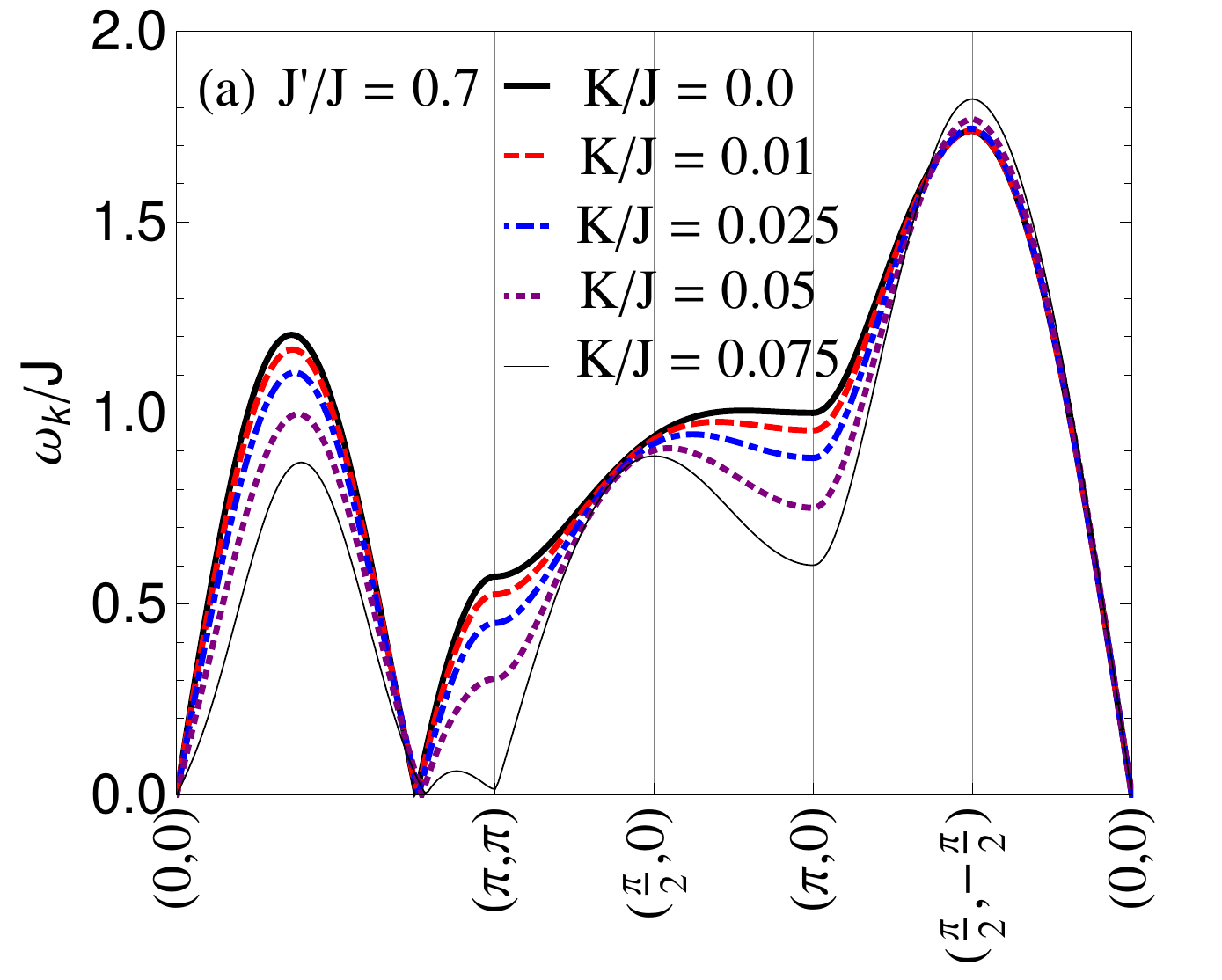}
 \includegraphics[width=0.66\columnwidth, clip]{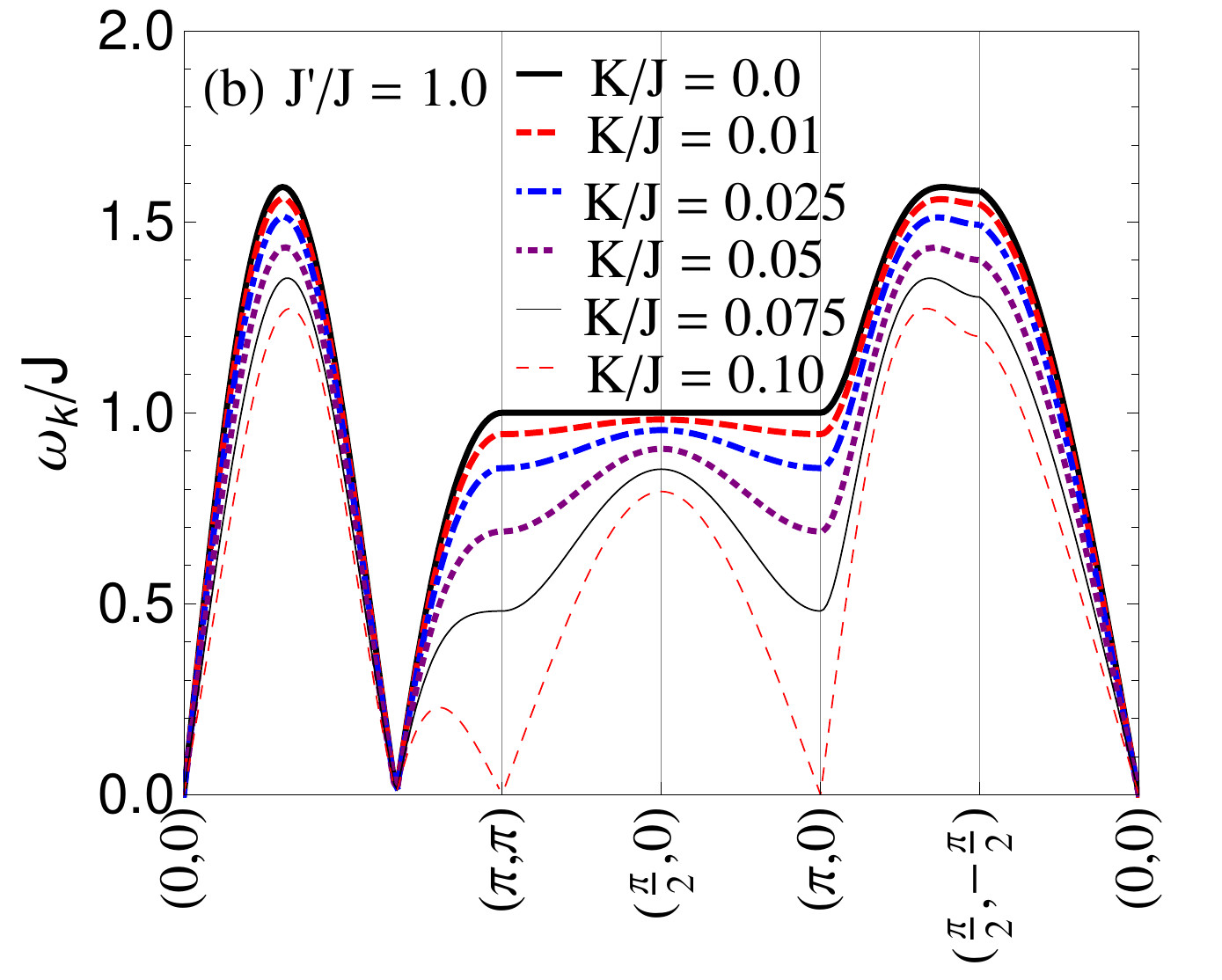}
\includegraphics[width=0.66\columnwidth, clip]{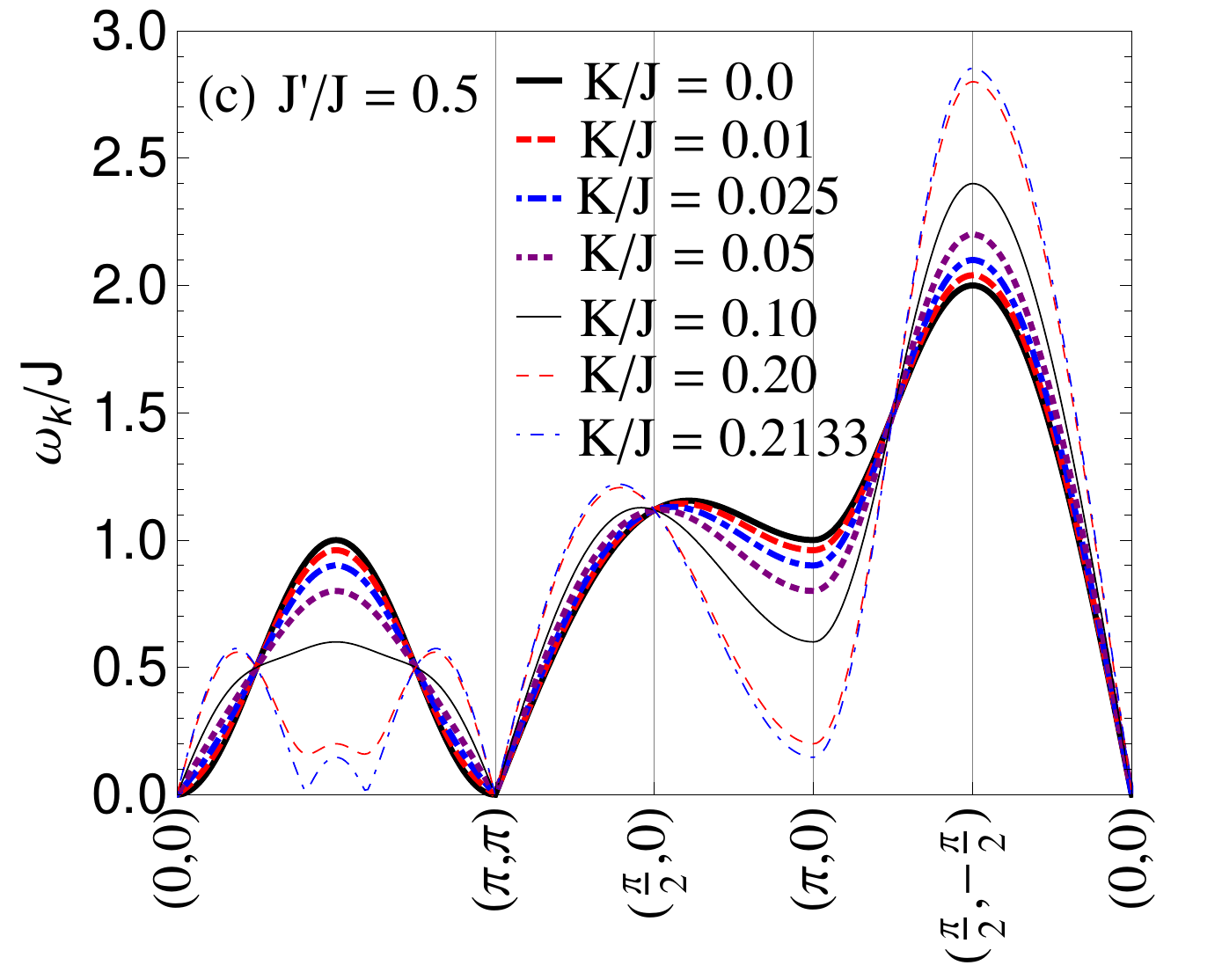}
    \end{center}
    \caption{(Color online) LSWT spectra for the spiral phase with $J'/J = 0.7$ (a) and $J'/J = 1.0$ (b) and for the N\'eel phase with $J'/J = 0.5$ (c). In (a) and (b) we mark spiral ordering vectors $q_{a}$ and $q_{b}$ with $q_{a} \approx 0.76\pi$ and $q_{b} =\frac{2}{3}\pi$. In all cases ring exchange increases the competition between the different classical phases, which causes a dramatically softening of the dispersion at the competing ordering wavevectors. Above the critical value of the ring exchange the dispersion becomes imaginary at these wavevectors - thus the competition between the different ordered phases is seen to be directly responsible for the quantum disordered phases.}
    \label{fig:dispersionring}
\end{figure*}

In Fig. \ref{fig:dispersionring} we present the spin-wave dispersion calculated from LSWT in both the N\'{e}el and spiral phases.  For the square lattice, $J'=0$, the spin-wave dispersion is independent of the ring exchange coupling $K$ in the multi-exchange model \eqref{eq:bigHamiltonian} considered here in contrast to related square lattice Heisenberg models \cite{Majumdar12, fnMSE}. 

In Fig. \ref{fig:dispersionring}(a) and (b) we plot the calculated spectra in the spiral phase for $J' = 0.7$ and $J'=J$.  In both cases one can clearly observe the expected Goldstone modes at ${\bf k} = {\bf 0}=(0,0)$ and ${\bf k} = {\bf Q}=(q, q)$. In the spiral phase increasing $K/J$ induces softenings at ${\bf k} = {\bm \pi} = (\pi, \pi)$ and ${\bf k} = (\pi, 0)$. For $J'/J = K'/K <1$ the mode softens most rapidly at ${\bf k}= \bm \pi$. For sufficiently large $K/J$ we find that $\omega_{\bm \pi}$ becomes imaginary (as $\omega_{\bm \pi}^2$ becomes negative) indicating that the competition with the N\'eel phase has destroyed the long range spiral order. For $J'/J = K'/K>1$ the mode softens most rapidly at ${\bf k}=(\pi, 0)$.  For sufficiently large $K/J$ we find that $\omega_{(\pi, 0)}$ becomes imaginary indicating that the competition with the collinear phase has destroyed the long range spiral order. At $J'/J = K'/K = 1$ (Fig.
  \ref{fig:dispersionring}(b)) both the N\'eel and collinear phases compete with the spiral phases (as one would suspect from the classical phase diagram, Fig. \ref{fig:phasering}(a)) and the dispersion becomes imaginary at ${\bf k} = \bm \pi$ and ${\bf k} = (\pi,0)$ simultaneously. A similar minimum at $(\pi,0)$ has been found from series expansions for the Heisenberg model on an ALT with no ring exchange due to recombination of particle-hole spinon pairs of momenta: $(\pi/2,\pi/2), (-\pi/2,-\pi/2)$ into magnons \cite{Zheng99, Fjaerestad07} in that case.

It is also interesting to note that for the most frustrated case, $J'=J$, the spiral order is more robust to the disordering effects induced by the ring exchange than for any other value of $J'/J$ even 
classically \cite{triangular}.  Furthermore, the strong geometrical frustration suppresses N\'{e}el and collinear phases thereby decreasing their ability to compete with spiral phase and drive an instability to the quantum spin liquid.

In the N\'{e}el phase (cf. Fig. \ref{fig:dispersionring} (c)) increasing $K/J$ leads to the softening of the mode at ${\bf k}=(\pi,0)$ and along the $\bf 0$-$\bm\pi$ direction. The later is more physically significant as it drives the N\'{e}el-spin liquid transition. For sufficiently large $K/J$ local minima in $\omega_{\bf k}$ emerge at ${\bf k} = {\bf k_N} =(k_N,k_N)$ and ${\bf k} = {\bm\pi} - {\bf k_N}$ with $k_N = \arccos([J' + 8K' - \sqrt{J'^2 - 24JK' + 16J'K' + 160K'^2}]/24K')$. As $K/J$ is further increased these minima deepen and eventually $\omega_{{\bf k_N}}=\omega_{{\bm\pi} - {\bf k_N}}$ becomes imaginary ($\omega_{\bf k_N}^2<0$). This is a clear indication that N\'eel order has become unstable due to competition with the spiral phase. Explicit calculation shows that long range spiral order with ${\bf Q}={\bf k_N}$ or $\bm\pi-\bf k_N$ is also unstable in this parameter regime. This is very different from the mechanism for the vanishing of long range order at the quantum critical point in the $K=0$ limit. 
 This is known \cite{Merino99,Trumper99} to be due to the vanishing of the spin-wave velocity along $\bf 0$-$\bm \pi$, which can be observed in Fig. \ref{fig:dispersionring}(c). In contrast for $K\ne0$ it is the competition between N\'eel $({\bf Q}=\bm\pi)$ and spiral $({\bf Q}={\bf k_N})$ order that destroys the long range order. 

We find that, in the parameter range overed by Fig. \ref{fig:phasering}, the collinear phase is always unstable to competition from other classically ordered phases. This means that there is always some point (or, typically, area) of the Brillouin zone for which $\omega_{\bf k}^2<0$. Therefore we conclude that competition with other classical phases means that the collinear phase is not stable in LSWT.

So far we have limited the discussion to the spin degrees of freedom only. However, in the materials of interest the charge degrees of freedom eventually become important and a Mott transition occurs under pressure.
For $J'=J$ (120$^\circ$) spiral order is found for $K/J\lesssim0.1$. To lowest order $U/t=\sqrt{20J/K}$, which would suggest that the spiral-spin liquid transition occurs at $U/t\simeq14$ which is in good agreement with the previous calculations of Motrunich \cite{Motrunich05} and Yang {\it et al}. \cite{Yang10} for the isotropic triangular lattice model. This is also close to the estimated value of the critical ratio of $U/t$ for the Mott transition on the triangular lattice \cite{Merino06}. This suggests that for $J'\sim J$ there is a direct transition from a spiral ordered Mott insulator to a metal as pressure is increased, which is believed to decrease $U/t$ \cite{Powell11, Kanoda10}. But, as sketched in the inset to Fig. \ref{fig:phasering}(b),  for smaller $J'/J$ one would find a spin liquid-metal phase transition. For yet smaller $J'/J$ these calculations predict a N\'eel-metal phase transition. Thus one would only expect to observe a Goldilocks spin liquid region
  where the quantum fluctuations due to geometrical frustration and ring exchange are sufficiently strong to suppress the N\'eel order, but not strong enough for the geometrical frustration to stabilize the spiral phase. Comparing the observed phase diagrams of the $\kappa$-(BEDT-TTF)${_2}X$ and $Y$[Pd(dmit)$_2$]$_2$ salts to this picture and taking into account the frustration ($J'/J$) estimated from first principles calculations \cite{Scriven12, Nakamura09, Kandpal09, Nakamura12, Tsumuraya13} one finds that this is exactly what is observed experimentally!

In this Letter we have shown that the competition between different long range order states creates a quantum disordered phase in the ATL Heisenberg model with ring exchange even at the semiclassical (LSWT) level. Our analysis suggests a spin liquid phase in a Goldilocks regime of frustration in which quantum fluctuations are sufficiently strong to induce competition between different classical orders without being  strong enough to stabilise the spiral phase via the reduction in competition with the other classical orders. Electronic structure calculations  \cite{Scriven12, Nakamura09, Kandpal09, Nakamura12, Tsumuraya13} show that the spin liquids $\kappa$-(BEDT-TTF)$_2$Cu$_2$(CN)$_3$  and Me$_3$EtSb[Pd(dmit)$_2$]$_2$ and the valence bond solid Me$_3$EtP[Pd(dmit)$_2$]$_2$ are all found in this Goldilocks regime. A future challenge is understanding ring exchange effects on two-dimensional metals close to the Mott transition which may lead to exotic non-Fermi liquid $d$-wave \cite{Fisher12} phases.

This work was funded in part by the Australian Research Council under the Discovery (DP130100757) and QEII (DP0878523) schemes.
J. M. acknowledges financial support from MINECO (MAT2012-37263-C02-01).


\begin{thebibliography}{99}
\bibitem{fn1}
Note in the convention followed in this Letter $J$ is a factor of two larger than in Ref. \onlinecite{Motrunich05}.
\bibitem{triangular} 

The effective two spin exchange couplings of  Hamiltonian (\ref{eq:bigHamiltonian}) are  $\tilde{J'}=J'+K+4K'$ and $\tilde{J}=J+2K+3K'$ implying that $\tilde{J'}=\tilde{J}$  for all $K/J$ when $J'/J=K'/K=1$.
\bibitem{fncpd} 

We also tested the columnar dimer and diagonal dimer phases with ordering vectors ${\bf Q} = (q - \pi, q)$ and  ${\bf Q} = (\pi - 2q, q)$ as well as the phase with ${\bf Q} = (q_{x}, q_{y})$, all of which were found to be classically unstable in the parameter region of interest
\bibitem{fnMSE}

Here we consider both the two-spin and four-spin terms of the four-spin permutation operators, whereas \cite{Majumdar12} considers only the four-spin terms. The spin dispersion is independent of ring exchange for $J' = 0$ in the multiexchange model (\ref{eq:bigHamiltonian}).
\end{thebibliography}
\end{document}